\documentclass[aps,prl,twocolumn,groupedaddress,showpacs]{revtex4}
\usepackage{amsmath}

\begin{document}

\title{Replica field theory and renormalization group for the Ising
       spin glass in an external magnetic field}

\author{T.~Temesv\'ari}
\email{temtam@helios.elte.hu}
\affiliation{HAS Research Group for Theoretical Physics,
E\"otv\"os University, P\'azm\'any P\'eter s\'et\'any 1/A,
H-1117 Budapest, Hungary}
\author{C.~\surname{De Dominicis}}
\affiliation{Service de Physique Th\'eorique,
        CEA Saclay, F-91191 Gif sur Yvette, France}

\date{\today}

\begin{abstract}
We use the generic replica symmetric cubic field-theory to study the
transition of short range Ising spin glasses in a magnetic field around
the upper critical dimension, $d=6$. A novel fixed-point is found, in
addition to the well-known zero magnetic field fixed-point, from the
application of the renormalization group. In the spin glass limit,
$n\to 0$, this fixed-point governs the critical behaviour of a class of
systems characterised by a single cubic interaction parameter. For this
universality class, the spin glass susceptibility diverges at criticality,
whereas the longitudinal mode remains massive. The third mode, the so-called
anomalous one, however, behaves unusually, having a jump at criticality.
The physical consequences of this unusual behaviour are discussed, and a
comparison with the conventional de Almeida-Thouless scenario presented.
\end{abstract}

\pacs{75.10.Nr, 05.10.Cc}

\maketitle

The mean-field theory of the Ising spin glass \cite{MePaVi} provided
an astounding complexity of equilibrium properties, showing up how
disorder and frustration may lead to an unusual thermodynamics. More
than two decades have elapsed since the
ultrametric solution of the Sherrington-Kirkpatrick \cite{SK} model
by Parisi was published in a series of papers (see \cite{MePaVi} for
the references), nevertheless consensus has not been reached about the
validity of the mean-field picture for finite-dimensional, short range
systems. The alternative scenario, the so-called droplet picture
\cite{BrMo86,FiHu86}, claims that the complex phase space structure is an
artifact of mean-field theory, the glassy state consists of two phases
related by the global inversion symmetry of the spins.

The investigation of the spin glass transition in an external magnetic
field may resolve the debate: The glassy transition along the
de Almeida-Thouless (AT) line \cite{AT} is a distinctive feature of mean-field
theory, whereas spin glass ordering is destroyed by any nonzero
magnetic field in the droplet model. Although a lot of numerical work
has been performed, no convincing evidence has emerged until now in favour
of either theory. An AT line was found in the four-dimensional case in
Refs.\ \cite{GrHe91,MaPaZu98}, whereas numerical results in three
dimensions were interpreted, although less convincingly, to support
mean-field-like behavior in \cite{CaPaPaSo90,GrHe91,MaPaZu00}.
On the other side, Ref.~\cite{HuFi91} interprets the simulation data of
\cite{CaPaPaSo90} as quite consistent with droplet theory, and an
analysis of the ground states in \cite{HoMa99} showed that the spin glass
phase of the three-dimensional model does not survive in any finite
magnetic field. More recently, however, an extensive study of the energy
landscape \cite{KrHoMaMaPa01} suggests that a nonzero critical field may
exist at zero temperature, separating the spin glass and paramagnetic phases.

In this letter, replica field theory, as an alternative to numerical
calculations, is used to attack the problem by extending the renormalization
group study of Ref.~\cite{BrRo}. Our starting point is the
Edwards-Anderson (EA) \cite{EA} model of $N$ Ising spins on a $d$-dimensional
hypercubic lattice, defined by the Hamiltonian
\begin{equation}\label{EAHamiltonian}
\mathcal{H}=-\sum_{\langle ij\rangle}J_{ij} s_is_j
-H\sum_i s_i
\end{equation}
where the first (second) summation is over all nearest neighbour
pairs $\langle ij\rangle$ (all lattice sites $i$),
respectively. 
$J_{ij}$ are independent, Gaussian distributed
random variables with mean zero and variance $\Delta ^2$,
and a homogeneous magnetic field $H$ has also been included.
The application
of the replica trick followed by a Hubbard-Stratonovich transformation
produces a {\em replica symmetric} field theory, with
the Lagrangean ${\mathcal L}_{\text{micr}}$ taking over
the role of
$\mathcal{H}$. In this replica field theory the fields depend on two
replica indices with the restriction $\phi^{\alpha\beta}=\phi^{\beta\alpha}$
and $\phi^{\alpha\alpha}\equiv 0$; hence we have $n(n-1)/2$ field components.
The replica number $n$ must go to zero to reproduce quenched averages; we
will argue, however, that it is necessary to keep it finite until the very
end of the calculations \footnote{The role of $n$ in the AT transition was
first raised by Kondor \cite{Ko83},
and elaborated in \cite{rscikk}. An interesting dynamical interpretation
of the replica number $n$ was put forward in \cite{PeCoSh93}.}.
As it is common in the theory of phase transitions,
the microscopic Lagrangean ${\mathcal L}_{\text{micr}}$ is replaced by an
effective one $\mathcal{L}=\mathcal{L}^{(2)}+\mathcal{L}^{(3)}$, obtained
by iterating the renormalization group until irrelevant operators can
be neglected. As a result of the replica trick, a {\em generic replica
symmetric} field theory follows, which can be best represented in terms
of operators invariant under the permutation of the $n$ replicas:
\begin{widetext}
\begin{equation}
\mathcal{L}^{(2)}=\frac{1}{2}\sum_{\mathbf p}\bigg[
\Big(\frac{1}{2} p^2+m_1\Big)\sum_{\alpha\beta}
\phi^{\alpha\beta}_{\mathbf p}\phi^{\alpha\beta}_{-\mathbf p}
+m_2\sum_{\alpha\beta\gamma}
\phi^{\alpha\gamma}_{\mathbf p}\phi^{\beta\gamma}_{-\mathbf p}+
m_3\sum_{\alpha\beta\gamma\delta}
\phi^{\alpha\beta}
_{\mathbf p}\phi^{\gamma\delta}_{-\mathbf p}
\bigg],\label{L2}
\end{equation}
\begin{equation}\label{L3}
\begin{aligned}
\mathcal{L}^{(3)}&=-\frac{1}{6\sqrt{N}}
\sideset{}{'}\sum_{\mathbf {p_1p_2p_3}}
\bigg[
w_1\sum_{\alpha\beta\gamma}\phi^{\alpha\beta}
_{\mathbf p_1}\phi^{\beta\gamma}_{\mathbf p_2}
\phi^{\gamma\alpha}_{\mathbf p_3}+w_2\sum_{\alpha\beta}\phi^{\alpha\beta}
_{\mathbf p_1}\phi^{\alpha\beta}
_{\mathbf p_2}\phi^{\alpha\beta}_{\mathbf p_3}
+w_3\sum_{\alpha\beta\gamma}\phi^{\alpha\beta}
_{\mathbf p_1}\phi^{\alpha\beta}
_{\mathbf p_2}\phi^{\alpha\gamma}_{\mathbf p_3}
+w_4\sum_{\alpha\beta\gamma\delta}\phi^{\alpha\beta}
_{\mathbf p_1}\phi^{\alpha\beta}
_{\mathbf p_2}\phi^{\gamma\delta}_{\mathbf p_3}\\
&+w_5\sum_{\alpha\beta\gamma\delta}\phi^{\alpha\beta}
_{\mathbf p_1}\phi^{\alpha\gamma}
_{\mathbf p_2}\phi^{\beta\delta}_{\mathbf p_3}
+w_6\sum_{\alpha\beta\gamma\delta}\phi^{\alpha\beta}
_{\mathbf p_1}\phi^{\alpha\gamma}
_{\mathbf p_2}\phi^{\alpha\delta}_{\mathbf p_3}
+w_7\sum_{\alpha\beta\gamma\delta\mu}\phi^{\alpha\gamma}
_{\mathbf p_1}\phi^{\beta\gamma}
_{\mathbf p_2}\phi^{\delta\mu}_{\mathbf p_3}
+w_8\sum_{\alpha\beta\gamma\delta\mu\nu}\phi^{\alpha\beta}
_{\mathbf p_1}\phi^{\gamma\delta}
_{\mathbf p_2}\phi^{\mu\nu}_{\mathbf p_3}
\bigg].
\end{aligned}
\end{equation}
\end{widetext}

The above field theoretical model is general enough to describe a
large variety of possible transitions from the high temperature,
replica symmetric phase, just below the upper critical dimension
$d=6$. The failure of Bray and Roberts \cite{BrRo} to detect a
fixed point corresponding to the conventional AT transition gave definite
support, from the analytic part, to the droplet theory. In this letter,
we present a novel fixed point characterized by a rather unusual property,
and propose
it as the relevant one for the {\em generic} replica symmetric phase,
i.e.\ that with a nonzero order parameter. We also put forward a
possible physical scenario whose validity is a prerequisite for that
fixed point to control the spin glass transition in a magnetic field.
Checking this, however, is out of the scope of this letter, and we leave
it for future work.

Even a leading order renormalization group calculation
is blocked by the difficulties arising from the numerous and complicated
replica summations, and from the fact that $\mathcal{L}^{(2)}$ is not in
a diagonalised form. In Ref.\ \cite{rscikk} we worked out a
transformation to a new set of bare parameters ($r_R$, $r_A$ and $r_L$ for
the masses, and $g_1,\ldots,g_8$ for the cubic couplings) rendering the
one-loop calculation feasible. As an illustration, we computed in
\cite{rscikk} the true masses $\Gamma_R$, $\Gamma_A$ and $\Gamma_L$
to one-loop order. To obtain the renormalized
cubic interaction, we calculated, via long but relatively
straightforward algebra, the triangle graph.
In order to be completely parallel to Ref.\ \cite{BrRo}, we chose the same
renormalization scheme of integrating out degrees of freedom in the
infinitesimal momentum shell between $e^{-dl}\,\Lambda$ and $\Lambda$,
where $\Lambda$ is the ultraviolet cutoff. It is obvious that
there is no sufficient space to present here the recursion relations in
their total generality, as worked out in \cite{Iveta}. The structure of
these RG equations is as follows:
\begin{align}
dr_i/dl&={\mathcal R}_i(r_{R},r_{A},r_{L};
g_1,\ldots,g_8)\qquad i=R,\,A\text{ or }L\,,\notag\\
dg_i/dl&={\mathcal G}_i(r_{R},r_{A},r_{L};
g_1,\ldots,g_8)\qquad i=1,\dots,8\,.\label{RGstructure}
\end{align}
The above set of equations must comprise two special cases, known from
the literature for some time, providing us with a good check:
\begin{enumerate}
\item The zero magnetic field case was studied in Ref.~\cite{Gr85} up
to $O(\epsilon^3)$, the Lagrangean corresponding to it ($r_{R}=r_{A}=
r_{L}\equiv r$, $w_1\equiv w$ and $w_i=0$ for $i=2,\dots,8$)
proves to be an invariant subspace of the set of Eqs.\ (\ref{RGstructure}),
expressing the higher symmetry this system possesses. The fixed point
attracts a critical line in $w-r$ space, which is totally massless
($\Gamma_R=\Gamma_A=\Gamma_L=0$), and, following the ideas in
\cite{rscikk}, we can identify this theory as the relevant
field-theoretical model for the spin glass transition from the paramagnet
to the generic replica symmetric phase with a nonzero order parameter.
The fixed-point value ${w^{*}}^2=-\frac{1}{n-2}\,\epsilon$, with
$\epsilon\equiv 6-d$, and the exponents $\eta$ and $\nu$ are
in complete agreement, up to first order in $\epsilon$
(see \cite{Iveta}), with the
results of Green \cite{Gr85} for $n=0$. As in mean-field theory, $n$ is
a rather innocent parameter around this fixed point.
\item 
The equations of Bray and Roberts \cite{BrRo} are reproduced by assuming
tentatively a critical surface with $\Gamma_R=0$, whereas $\Gamma_A$ and
$\Gamma_L$ finite. At some hypothetical fixed-point, the bare anomalous and
longitudinal masses are infinite, resulting in a pair of recursion relations
for the replicon-like couplings $g_1$ and $g_2$:
\begin{equation}\label{Rmassless}
dg_i/dl=\bar{\bar{{\mathcal G}}}_i(g_1,g_2)\qquad i=1,2\,.
\end{equation}
Using the relations $g_1=w_1$ and $g_2=2w_2$ \cite{rscikk}, we arrive at the
RG recursions whose physically relevant fixed-point was searched for in vain
by  Bray and Roberts \cite{BrRo}.
\end{enumerate}

A new theory, invariant under renormalization to $O(\epsilon)$ order, emerges
if the condition of degeneracy between the longitudinal (L) and
anomalous (A) modes is removed. The physical picture behind this may be
the following: The masses characterizing a critical manifold are
observable through correlation functions. For the original lattice
system of Eq.\ (\ref{EAHamiltonian}), three distinct spin glass
correlation functions can be defined (see e.g.\ in \cite{BrMo86}),
their zero momentum limits will be denoted by $G_1$, $G_2$ and
$G_3$.
The transition in a field can be characterized
by the divergence of the spin glass susceptibility, $\chi_{\text{SG}}=
G_1-2G_2+G_3=\Gamma_R^{-1}$, manifesting itself in the criticality of
the replicon (R) mode, while the longitudinal mode,
$G_1-4G_2+3G_3=\Gamma_L^{-1}$, remains massive. As for the anomalous one,
we have the exact relation for $n$ small but finite:
\begin{equation}\label{Abasic}
\Gamma_A=\frac{\Gamma_L}{1-n\Gamma_L(G_2-\frac{3}{2}G_3)}.
\end{equation}
In mean-field theory, the combination $G_2-\frac{3}{2}G_3$ is also analytical
along the AT-line, leading to the following ratios for the {\em leading}
singularities: $G_1:G_2:G_3=1:1/2:1/3$. Dimension-dependent subleading
singularities of the form $\sim t^{-\mu}$ occur in finite dimensions,
$t$ being the reduced temperature, and an analysis of the perturbation
expansion of the propagators shows that for the next-to-leading term, $\mu$
is equal to $4-d/2$, $d>6$. As a result, a jump of the anomalous mass may
develop below $d=8$ \footnote{The idea that a qualitative change occurs in
the properties of the AT transition at $d=8$ was first suggested in
Ref.\ \cite{GrMoBr83}.}, provided the longitudinal mode remains massive. From
Eq.\ (\ref{Abasic}) it follows:
\begin{equation}\label{cases}
\Gamma_A\sim
\begin{cases}\Gamma_L& \text{if $n\to 0$ first},\\
             \frac{1}{n}t^{\mu}& \text{if $t\to 0$ first}.
\end{cases}
\end{equation}

From the assumptions $\Gamma_L$ finite and $\mu>0$, and using (\ref{Abasic}),
a jump in the anomalous mass necessarily follows in the limit $n\to 0$
even below $d=6$.
Whether or not these assumptions are correct, can be tested by the RG
equations (\ref{RGstructure}). To detect this behavior, we now search for
a nontrivial fixed point with $\Gamma_R=\Gamma_A=0$ and $\Gamma_L=\infty$.
It is obvious from Eq.\ (\ref{cases}) that $n$ is a crucial parameter now,
and it must be finite when computing the RG flows, setting it to $0$ only
at the very end of the calculation
\footnote{You can find the fixed-point (\ref{specialfixed-point}) even
if you start with $n=0$, the RG flows, however, are ill-behaved around it:
the Hessian is a non-diagonalisable matrix then, and the fixed-point
itself is not an isolated one, but sitting on a line of {\em equivalent}
fixed-points. As a matter of fact, the limit $n\to 0$ and the iteration of
the RG is {\em not} interchangeable around that fixed-point. Just these
mathematical findings led us to throw away the condition of degeneracy
between the L and A modes. From the physical point of view, the assumption
of the divergence of the second term in the denominator of Eq.\
(\ref{Abasic}), for nonzero $n$, is the crucial deviation from conventional
thinking about the AT transition. It was implicit in Ref.\ \cite
{BrRo} that the replicon combination is the only correlation function
divergent at criticality.}.
The longitudinal bare mass is kept at its
infinite fixed-point value, longitudinal-like couplings ($g_4$,
$g_7$ and $g_8$; see Eq.\ (48) of Ref.\ \cite{rscikk}) decouple from the rest
of Eq.\ (\ref{RGstructure}). A massless renormalization scheme (massless with
respect to the replicon (R) and anomalous (A) masses) can be deduced from
the remaining part of (\ref{RGstructure}):
\begin{equation}\label{RAmassless}
dg_i/dl=\bar{{\mathcal G}}_i(g_1,g_2,g_3,g_5,g_6)\qquad
i=1,2,3,5,6\,.
\end{equation}
In our one-loop calculation, $\bar{{\mathcal G}}_i$ is a cubic polynomial
of its variables with coefficients which are rational functions of $n$.
It turns out that $\bar{{\mathcal G}}_3$ and $\bar{{\mathcal G}}_6$ are
always zero whenever $g_3=g_6=0$ \cite{Iveta}, indicating that the
3-dimensional manifold so defined is an invariant subspace of the RG equations
(\ref{RAmassless}). Instead of presenting long and cumbersome formulae for
generic $n$, we display the recursion relations for $g_1$, $g_2$ and
$g_5$ only in the $n\to 0$ limit:
\begin{widetext}
\begin{subequations}
\begin{align}
\label{g1}\frac{dg_1}{dl}&=\frac{1}{2}(\epsilon-3\,\eta_R)g_1+
14\,g_1^3-18\,g_1^2g_2+\frac{9}{2}\,g_1g_2^2+\frac{1}{8}\,g_2^3-
8\,g_5^3\,,\displaybreak[0]\\
\label{g2}\frac{dg_2}{dl}&=\frac{1}{2}(\epsilon-3\,\eta_R)g_2+
24\,g_1^2g_2-30\,g_1g_2^2+\frac{17}{2}\,g_2^3\,,
\displaybreak[0]\\
\label{g5}\frac{dg_5}{dl}&=\frac{1}{2}(\epsilon-\eta_R-2\,\eta_A)g_5-
8\,g_1g_5^2+8\,g_2g_5^2+8\,g_5^3\,;\displaybreak[0]\\[8pt]
\label{etaR}\eta_R&\equiv \frac{1}{3}\,(4\,g_1^2-8\,g_1g_2+
\frac{11}{4}\,g_2^2+4\,g_5^2)\,.
\end{align}
\end{subequations}
\end{widetext}
($\eta_A$ above is proportional to $n$, i.e.\ it is zero here. It is shown
only to display the generic structure of the RG equations.)

We found a novel non-trivial fixed-point from Eqs.\ (\ref{g1}), (\ref{g2})
and (\ref{g5}):
\begin{equation}\label{specialfixed-point}
g_1^*=\sqrt{\epsilon}/2,\qquad g_2^*=\sqrt{\epsilon},
\qquad g_5^*=-\sqrt{\epsilon}/4.
\end{equation}
The existence of this fixed-point is due to the term $g_5^3$ and $g_5^2$
in Eqs.\ (\ref{g1}) and (\ref{etaR}), respectively; omitting them, we just
get back the Bray-Roberts equations (\ref{Rmassless}). It is remarkable that,
in a sense, we have found in Eqs.\ (\ref{g1}), (\ref{g2}), (\ref{g5}) and
(\ref{etaR}) a generalization of the RG theory put forward
in Ref.\ \cite{BrRo}. We must notice, however, that beside the replicon mode,
the anomalous one is also critical on the manifold attracted by the
fixed-point (\ref{specialfixed-point}).

The most striking feature of Eqs.\ (\ref{g1}), (\ref{g2}) and (\ref{g5})
is that they coincide for $g_1=\bar w$, $g_2=2\bar w$ and $g_5=-\bar w/2$
providing the single parameter RG equation for $\bar w$:
\[\frac{d\bar{w}}{dl}=\frac{\epsilon}{2}\bar{w}-2\bar{w}^3.
\]
Translating this to the language of the $w$ couplings in Eq.\ (\ref{L3}),
$w_1=w_2=w_6=\bar{w}$, and all the other $w$'s are zero. The system with
the single cubic operator
\begin{equation}\label{invariant}
\bar{w}\,\bigg[\sum_{\alpha\beta\gamma}\phi^{\alpha\beta}
_{\mathbf p_1}\phi^{\beta\gamma}_{\mathbf p_2}
\phi^{\gamma\alpha}_{\mathbf p_3}+
\sum_{\alpha\beta}\phi^{\alpha\beta}
_{\mathbf p_1}\phi^{\alpha\beta}
_{\mathbf p_2}\phi^{\alpha\beta}_{\mathbf p_3}+
\sum_{\alpha\beta\gamma\delta}\phi^{\alpha\beta}
_{\mathbf p_1}\phi^{\alpha\gamma}
_{\mathbf p_2}\phi^{\alpha\delta}_{\mathbf p_3}\bigg]
\end{equation}
rescales under RG, evolving into the fixed point $\bar{w}^*=\sqrt{\epsilon}
/2$, where the corresponding eigenvalue is $\lambda_{\bar{w}}=-\epsilon$.
This result can be compared with the zero-field case, where the similar
rescaling property of the system expressed a higher symmetry than the
permutation
invariance of the $n$ replicas. There is one important distinction we must
notice, however: the rescaling behavior of the zero-field Lagrangean
under iteration is
independent of $n$, whereas it develops only in the spin glass limit $n\to 0$
for the system with the cubic coupling in (\ref{invariant}).

Including the masses into the RG scheme, it can be easily checked that the
condition $r_R=r_A\equiv \bar{r}$ is preserved under iteration. Using
results from Ref.~\cite{rscikk}, namely Eqs.\ (22)--(24) and (28)--(30),
$m_1=\bar{r}/2$ and $m_2=0$ follow then, while $m_3$ is infinite,
inducing the freezing-out of the longitudinal component of
$\phi^{\alpha\beta}$. The quadratic operator in the brackets of
Eq.\ \ref{L2} reduces to the simple replicon-like invariant
\begin{equation}\label{massinvariant}
(p^2+\bar{r})\sum_{\alpha<\beta}
\phi^{\alpha\beta}_{\mathbf p}\phi^{\alpha\beta}_{-\mathbf p},
\end{equation}
although $\phi^{\alpha\beta}$ has now an anomalous component too.

We can deduce critical indices belonging to this new fixed point, and
we display them here for completeness:
\begin{alignat*}{2}
\eta_R&=O(\epsilon^2),&\qquad\lambda_R&=\nu^{-1}_R=2-\frac{\epsilon}{2}+
O(\epsilon^2);\\
\eta_A&=O(\epsilon^2),&\qquad\lambda_A&=\nu^{-1}_A=2+O(\epsilon^2).
\end{alignat*}
To connect these to usual exponents like that of the spin glass
susceptibility, $\chi_{\text{SG}}\sim t^{-\gamma}$, is not trivial now
due to the coexistence of two critical masses, and needs further study.

We propose the simple model of Eqs.\ (\ref{massinvariant}) and
(\ref{invariant}) as a candidate for studying the replica symmetry breaking
(RSB) transition from the replica symmetric phase with a nonzero order
parameter; the spin glass transition in an external field belongs to this
class. Nevertheless, it is difficult to find evidence for this. The check
by testing the crossover, {\em from\/} a $\mathcal{L}_{micr}$ in the
vicinity of the zero-field critical point {\em  to\/} the new fixed point,
is blocked by the large distance in this
huge parameter space, and by the evolution of the longitudinal mass from
the near-zero value to infinity. It is also obvious that the irrelevant
operators present in $\mathcal{L}_{micr}$ influence this crossover, rendering
this check very difficult. To by-pass this problem, it is tempting to imagine
an alternative scenario, viz.\ the existence of the replica symmetric phase
with nonzero order parameter even in zero field. This two-step process from
the paramagnet to the RSB phase is present in mean-field, but only for
finite, albeit infinitesimal, $n$ \cite{rscikk}. If this scenario occured
in low enough dimensions even for $n=0$, the crossover from one type of
transition to the other would disappear. In this case, the intermediate
replica symmetric phase with $Q\not=0$ would have the resemblance to a
droplet-like phase. It is clearly necessary to perform further
investigations, mainly a higher order calculation and numerical
investigations.

In conclusion, we must stress that the theory we have put forward for the
RSB transition is qualitatively different from the AT transition of
mean-field theory. We argued that the change occurs at $d=8$, below which
the relevant Gaussian theory is that with zero replicon and anomalous
masses, while infinite longitudinal one. It is this Gaussian fixed-point
which gives birth to the nontrivial one we found, governing the RSB
transition below $d=6$. We can speculate that these qualitative
differences may affect the glassy phase too, resulting in a more general
RSB scheme than the ultrametric one of mean-field theory.

\begin{acknowledgments}
Helpful discussions with E.~Brezin, L.~Sasv\'ari and I.~Kondor are highly
appreciated. We are especially grateful to Imre Kondor for a critical
reading of the manuscript.
This work has been supported by the Hungarian Science Fund (OTKA),
Grant No.~T032424.
\end{acknowledgments}


\end{document}